\documentclass[fleqn,10pt]{wlscirep}
\usepackage[utf8]{inputenc}
\usepackage[T1]{fontenc}
\usepackage{float}

\title{Free log-likelihood as an unbiased metric for coherent diffraction imaging}

\author[1,2,*]{Vincent Favre-Nicolin}
\author[1]{Steven Leake}
\author[1]{Yuriy Chushkin}
\affil[1]{ESRF, The European Synchrotron, 71 Avenue des Martyrs, 38000 Grenoble, France}
\affil[2]{Univ. Grenoble Alpes, Grenoble, France}
\affil[*]{favre@esrf.fr}

\begin{abstract}
Coherent Diffraction Imaging (CDI), a technique where an object is reconstructed from a single (2D or 3D) diffraction pattern, recovers the lost diffraction phases without \textit{a priori} knowledge of the extent (support) of the object. The uncertainty of the object support can lead to over-fitting and prevents an unambiguous metric evaluation of solutions. We propose to use a 'free' log-likelihood indicator, where a small percentage of points are masked from the reconstruction algorithms, as an unbiased metric to evaluate the validity of computed solutions, independent of the sample studied. We also show how a set of solutions can be analysed through an eigen-decomposition to yield a better estimate of the real object. Example analysis on experimental data is presented both for a test pattern dataset, and the diffraction pattern from a live cyanobacteria cell. The method allows the validation of reconstructions on a wide range of materials (hard condensed or biological), and should be particularly relevant for 4th generation synchrotrons and X-ray free electron lasers, where large, high-throughput datasets require a method for unsupervised data evaluation.
\end{abstract}
\begin{document}

\flushbottom
\maketitle
\thispagestyle{empty}

\section*{\label{sec:intro}Introduction}
Coherent Diffraction Imaging (CDI) is a technique that exploits the coherence properties of a light source, for instance, synchrotron generated X-ray beams, to reconstruct two- or three-dimensional objects from their diffraction pattern alone \cite{sayre_extendibility_1998, miao_extending_1999, miao_possible_2000, miao_approach_2001, marchesini_x-ray_2003}. Such an approach can also be used with soft X-ray sources \cite{sandberg_lensless_2007} and coherent electron beams \cite{zuo_atomic_2003, huang_coordination-dependent_2008} and when employed in the Bragg geometry yields quantitative strain information \cite{robinson_reconstruction_2001, williams_three-dimensional_2003, pfeifer_three-dimensional_2006, favre-nicolin_analysis_2010, robinson_coherent_2009}. CDI has been successfully used on a wide range of samples, from single cells \cite{shapiro_biological_2005} to inorganic particles \cite{chushkin_three-dimensional_2014, beuvier_x-ray_2019}, with applications exploiting the temporal properties of X-ray Free Electron Lasers to viruses \cite{seibert_single_2011} and time-resolved strain analysis \cite{clark_ultrafast_2013}.

As CDI is based on the measurement of the far-field diffraction pattern of a single object, the reconstruction is only possible if the diffraction pattern is recorded at a spacing finer than the Nyquist frequency  (this condition is called \textit{oversampling})\cite{sayre_implications_1952,sayre_extendibility_1998, miao_possible_2000}. This is easily done experimentally if the sample size can be estimated, and a variety of algorithms (Error Reduction (ER), Hybrid Input-Output (HIO), Relaxed Averaged Alternating Reflectors (RAAR), Charge Flipping (CF) etc.) can be used to phase the diffraction pattern and reconstruct the object \cite{marchesini_x-ray_2003, marchesini_unified_2007, fienup_phase_2013}.

However the weakness of CDI lies in the absence of reliable figures of merit to assess the quality of the reconstructed objects. In principle, it is easy to define a figure of merit by comparing the observed diffraction pattern to the calculated one. But as the diffraction pattern is oversampled and the actual size and shape of the object is unknown, it is easy to create incorrect solutions which involve an object size larger than the real one (i.e. with many extra free parameters), and thus yield a better figure of merit by over-fitting.

In this article, we propose a free log-likelihood as an objective figure of merit that outperforms those that exist in the literature. Then we show how it can be applied to evaluate the solutions and combine them to obtain the final optimal reconstruction.

All the data and the python notebooks used to generate the figures in this article are available from \cite{favre-nicolin_free_2019}.

\section*{\label{sec:figures_of_merit}Figures of merit}
A number of figures of merit have been used in the literature, working either in the object or Fourier-space domain. A list of the most used figures of merit is shown in table \ref{tab:fig_merit}. Note that the most quantitative approach to a reliable figure of merit was introduced for Ptychography \cite{thibault_maximum-likelihood_2012}, using a likelihood based analysis that considers Poisson noise.

\begin{table*}[ht]
    \centering
    \begin{tabular}{|c|c|c|c|c|c|c|}
    
        \hline
        Figure of merit & tight (1b) & +2 pixels (1c) &large (1d)&eigen-10 (3a)&average-10 (3b) & eigen-4 \\
        \hline
        $nb_{support}$ & 4353 & 10460 & 17616 & - & - & - \\
        \hline
        $E_o^2 = \sum\limits_{i \not\in \Omega } \lvert\rho_i\rvert^2 / \sum\limits_{i} \lvert\rho_i\rvert^2$ & 4.9e-3& 3.3e-3& 4.5e-3& 1.4e-2& 3e-2 & 6e-3 \\
        \hline
        $E_F^2 = \sum \lvert F^{calc}_i-F^{obs}_i\rvert^2 / \sum \lvert F^{obs}_i \rvert^2$ & 0.8559 & 0.8557 & 0.8558& 0.8590 & 0.8610& 0.8570\\
        \hline
        $LLK = -\frac{1}{N} \sum\limits_{i}\log \frac{(I^{calc}_i)^{I^{obs}_i}}{I^{obs}_i!}e^{-I^{calc}_i}$ & 68 & 46&  45 & 100 & 148 & 82 \\
        \hline
        $LLK_{free}$ & 83 & 122& 323& 98 & 149 & 80\\
        \hline
    \end{tabular}
    \caption{Figures of merit for CDI analysis: $E_o^2$ is the object-domain error \cite{miao_approach_2001, fienup_phase_2013}, where $\Omega$ denotes the object support (i.e. the area or volume where the object lies), and $\rho_i$ is the density inside the object. $E_F^2$ is the Fourier-space domain error \cite{miao_approach_2001, chapman_high-resolution_2006, fienup_phase_2013} comparing calculated and observed amplitudes (this formula can also be used based on intensities). $LLK$ is the Poisson log-likelihood\cite{thibault_maximum-likelihood_2012}, where $I^{obs}_i$  and $I^{calc}_i$ are, respectively, the observed and calculated intensity at pixel $i$ of the diffraction data. $LLK_{free}$ is the Poisson log-likelihood computed only over a 'free' set of pixels. The figures correspond either to the three solutions shown in Fig.\ref{fig:fig1}, or to the eigen- and average solutions shown in Fig.\ref{fig:fig_modes}. Only the $LLK_{free}$ allows to correctly discriminate between individual reconstructions.}
    \label{tab:fig_merit}
\end{table*}

\begin{figure}[ht]
    \centering
    \includegraphics[width=0.5\columnwidth]{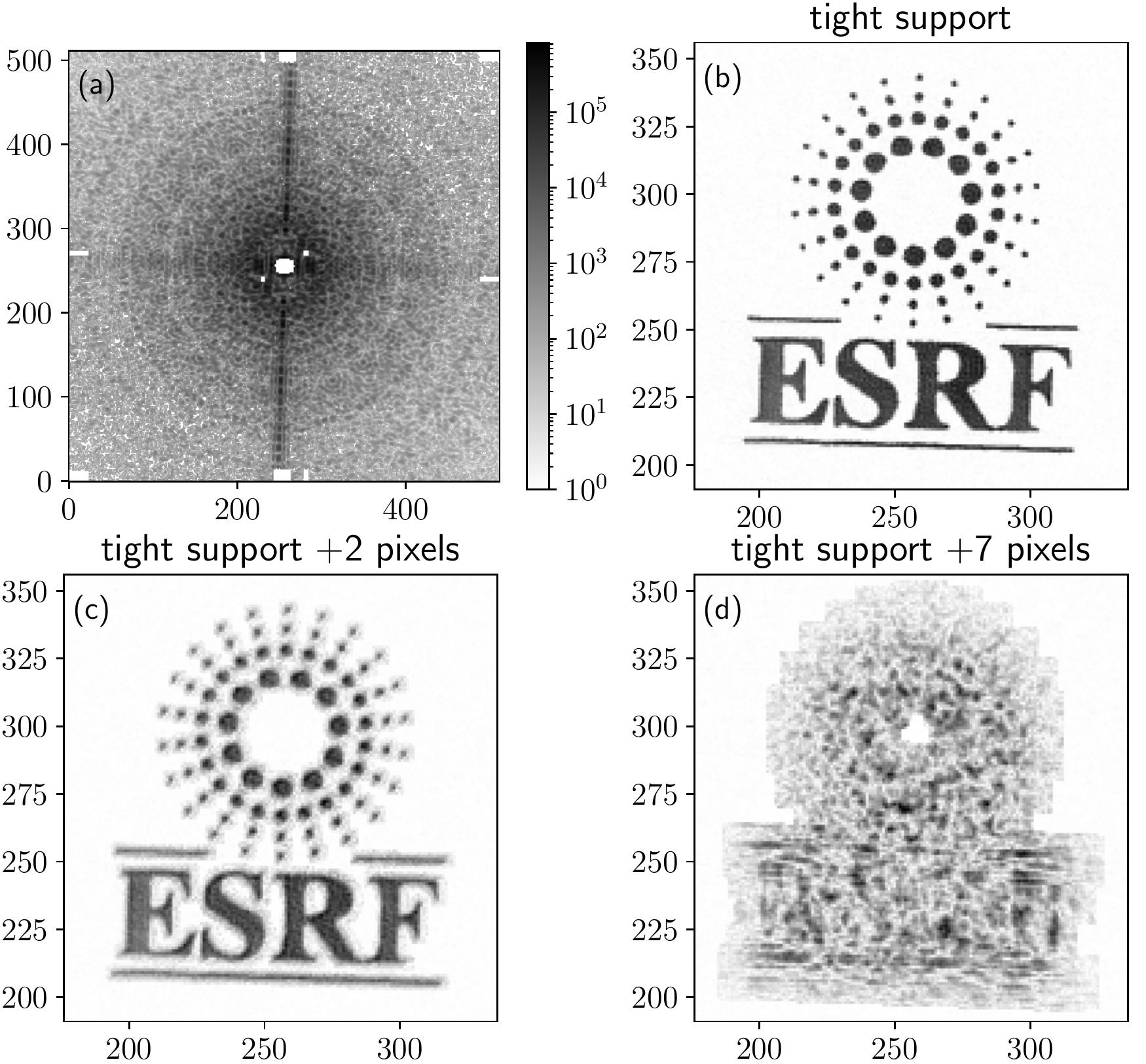}
    \caption{(a) Coherent X-ray diffraction pattern of the ESRF logo. Examples of the reconstructed logo obtained using (b) a tight support, (c) a tight support expanded by two pixels (producing more noise around the object), and (d) a tight support expanded by seven pixels. For each reconstruction, the average Poisson log-likelihood is reported in table \ref{tab:fig_merit}, as well as the free log-likelihood - the former does not allow to discriminate solutions, and allows clearly diminished results like (d) to have lower log-likelihood, whereas the free log-likelihood clearly discriminates them.}
    \label{fig:fig1}
\end{figure}

In order to evaluate how discriminating these figures are, we used a 2D diffraction dataset, from an ESRF logo sample,  recorded at the ID10 beamline \cite{chushkin_three-dimensional_2014} (see \cite{latychevskaia_imaging_2015} for experimental details). A center of symmetry was applied to the diffraction pattern in order to reduce the number of unobserved pixels behind the beamstop (in the case of a homogeneous object with a constant thickness, where the projected electronic density is constant i.e., the object is real up to a constant phase factor, the diffracted intensity (Fourier Transform) is centro-symmetric).  In Fig.\ref{fig:fig1} the diffraction data and several reconstructed solutions are shown, which were obtained by starting from a random object with a fixed support, 400 cycles of Hybrid Input-Output followed by 200 cycles of Error Reduction. Note that the actual algorithm used here is irrelevant, as any global optimisation scheme will produce a distribution of solutions, and the aim of this article is to provide a way to assess their quality. We deliberately did not impose positivity to have a large enough range of solutions to evaluate. 

The three solutions were generated using (b) a tight support (4353 pixels), and loose supports that radially expand the tight support by (c) two pixels, or (d) seven pixels.. As is clearly seen in Fig. \ref{fig:fig1}, the best result is obtained with a tight support, and quickly degrades for larger supports.
As can be seen in table \ref{tab:fig_merit}, none of the figures of merit are discriminating: while the solution in Fig. \ref{fig:fig1}b) obtained from a tight support is obviously the best and corresponds to the scanned sample \cite{latychevskaia_imaging_2015}, both the object-domain error and the Fourier-based metric $E_F^2$ show little difference between the solutions, and are higher for the correct tight support solution. This is clearer for the $LLK$, which is worse for the tight solution (b) even though it is the best result. The main reason for these conflicting results is that the more points there are in the support, the more free parameters the algorithm can use to better fit the diffraction data - Fig. \ref{fig:fig1}d) shows a clear case of over-fitting.

From this example one concludes that it is essential to simultaneously achieve both: a good fit between the calculated and observed diffraction pattern, and a tight support around the object. This has long been known and efficient algorithms, such as the shrink-wrap approach \cite{marchesini_x-ray_2003}, exist to produce a tight support which will normally yield a unique solution in two or three dimensions \cite{devaney_uniqueness_1978, crimmins_ambiguity_1981, bates_fourier_1982, hayes_reconstruction_1982, bates_uniqueness_1984, seldin_numerical_1990}.

Other constraints or figures of merit can be used based on physical properties, but they rely strongly on \textit{a priori} knowledge of the object and thus are limited in application. For example, positivity of the reconstructed object (e.g. when operating in the small-angle regime for a thin object, or in the Bragg regime for an unstrained nanocrystal), or enforcing phase, density or other \textit{ad hoc} constraints\cite{ulvestad_identifying_2017}.

The lack of an unambiguous evaluation metric often leads to the need to visually inspect, evaluate and select solutions, which is very slow, and hence unpractical. As brighter X-ray sources pave the way for serial CDI experiments with larger data throughput \cite{schulli_x-ray_2018, bjorling_coherent_2019}, the need, for a discriminating figure of merit to sort the computed solutions without requiring a visual inspection, is paramount.

\section*{\label{sec:free_llk}free log-likelihood}

The risk of misleading figures of merit due to over-fitting also exists e.g. in macro-molecular crystallography, where the large number of parameters (the atomic positions) must be refined against the available diffraction data: in order to avoid over-fitting, a free R-factor was introduced \cite{brunger_free_1992, tickle_rfree_1998,tickle_rfree_2000} and has since been used by the community. This consists of setting aside a small percentage of diffraction data (the 'free' set) and refining the structure against the remaining 'working set' of data. The $R_{free}$ is then evaluated only by comparing the calculated diffraction data against the 'free' set, producing an un-biased figure of merit. This approach is more generally known as jack-knifing \cite{quenouille_problems_1949, efron_jackknife_1981}, and was developed for unbiased statistical evaluation.

This approach is even more appealing for CDI because of the large uncertainty of the support area (contrary to refinement in crystallography where the atomic sequence or chemical formula is usually known). We have implemented this in the PyNX accelerated coherent imaging toolkit \cite{favre-nicolin_pynx_2010}, which can be used for CDI analysis:
\begin{itemize}
    \item $\approx$5\% of the observed diffraction data is set aside in a 'free' set of pixels with pixels, which are grouped in islands of radius 3 pixels (volumes in 3D) - to make sure that correlations between neighbouring pixels does not create a strong relationship between the working set and the 'free' set. This is needed because the diffraction data is oversampled \cite{sayre_extendibility_1998, miao_oversampling_2000}, and thus neighbouring pixels are not completely independent. Note that the proposed size of the islands corresponds to a typical oversampling ratio in CDI experiments, but could be tuned according for each experiment, even each direction. The effect of the island size can be seen in figures available as supplementary information.
    \item The free pixels are randomly located in the dataset, only excluding the center of the diffraction pattern (5\% of the maximum radius): this area can include a large number of photons, and masking those can hinder the initial estimate of the object.
    \item when performing the usual projection algorithms \cite{marchesini_unified_2007}, in the Fourier update step (replacing the calculated amplitudes with the observed ones while keeping the calculated phases), pixels in the 'free' set are considered masked and keep their calculated complex value.
\end{itemize}

In the following we will only report the free Poisson log-likelihood $LLK_{free}$ figure of merit, because the photon counting properties of modern X-ray detectors theoretically make Poisson log-likelihood the natural choice. However, we would like to point out that the ensuing discussion and examples would be identical with any other noise model or Fourier based metric.

As is shown in Table. \ref{tab:fig_merit}, the free log-likelihood correctly discriminates between the three solutions (tight support, +2 pixels, +7 pixels).

\begin{figure}[H]
    \centering
    \includegraphics[width=0.55\columnwidth]{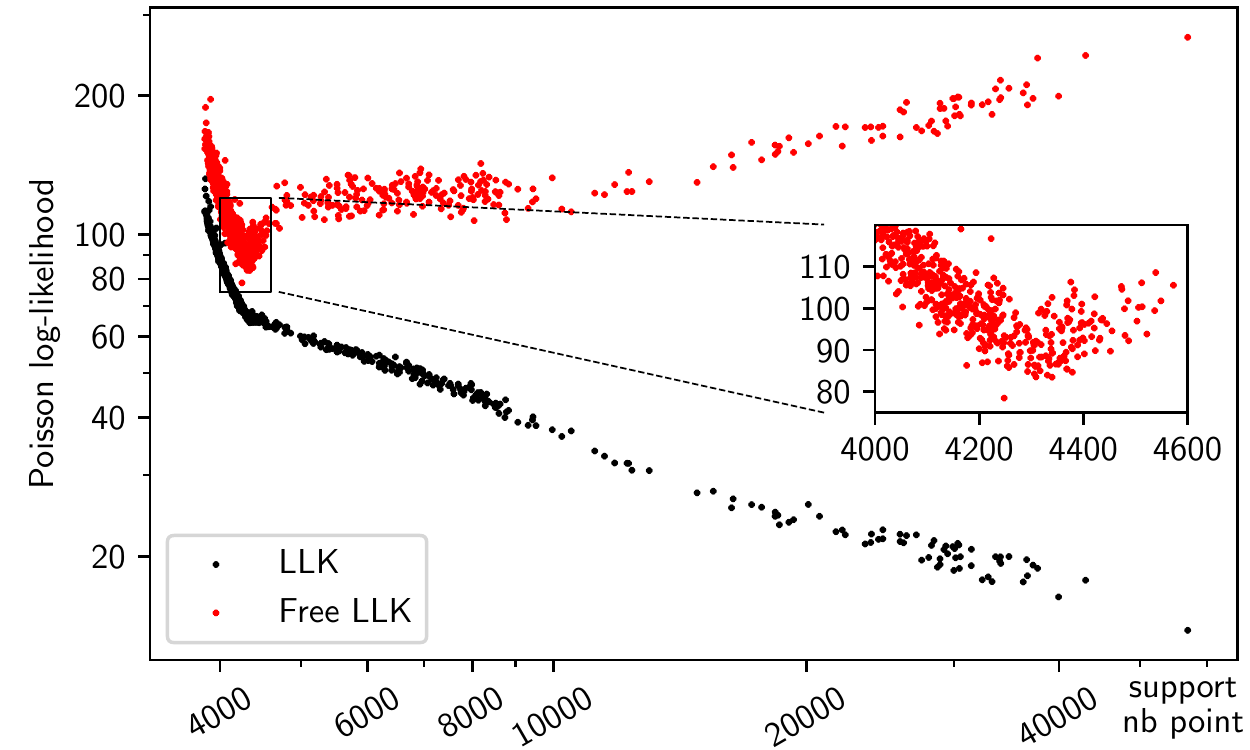}
    \caption{Scatter plot of log-likelihood (black) and free log-likelihood (red) vs the number of points in the support, obtained by generating 1000 solutions with random starting objects and different parameters. While the overall log-likelihood decreases with the number of points in the final support, the free log-likelihood points to an optimal support size around 4300 points. Note that a different, random set of 'free' pixels was used for each solution.}
    \label{fig:fig_llk_evolution}
\end{figure}

We conducted a more systematic study using the same dataset as for Fig.\ref{fig:fig1} and performed 1000 optimisations using a similar approach: starting from a random object with an initial tight support expanded by a radius of 7 pixels, then performing 400 HIO cycles followed by 200 ER cycles, updating the support every 20 cycles, with a support threshold\cite{marchesini_x-ray_2003} randomly chosen between 0.25 to 0.4. Note that we did not impose positivity: that and the random threshold values leads to a wide range of solutions for statistical purposes. With a positivity constraint, most solutions would have been much closer to the optimal one. Both $LLK$ and $LLK_{free}$ are plotted against the final number of pixels in the support for all solutions in Fig. \ref{fig:fig_llk_evolution}. In this graph the normal $LLK$ (measured against the 'working' set) monotonically decreases as the number of points in the support increases, even if a kink is clearly visible in the curve. $LLK_{free}$ however, displays a clear minimum around 4300 pixels, which corresponds to an optimal solution similar to that in Fig. \ref{fig:fig1}b.

While this demonstrates the capability of $LLK_{free}$ to serve as an unbiased figure of merit, we have identified two limitations. First, nothing prevents an incorrect model to have a low $LLK_{free}$ by chance - however as is shown in Fig. \ref{fig:fig_llk_evolution} with 1000 generated solutions, it is statistically improbable. Second, $LLK_{free}$ is not an \textit{absolute} figure of merit, and a single value cannot indicate the validity of the solution, as it is dependent on the counting statistics (e.g. a dataset with many zero-valued points will generate both a low $LLK$ and a low $LLK_{free}$). Finding an optimal solution relies on (i) generating a number (typically at least 20) of solutions and then (ii) selecting the ones with the lowest $LLK_{free}$, for the same set of free pixels. A more complete validation requires a statistical analysis of the best solutions, as will now be explained.

\section*{\label{sec:solution_analysis}Eigen- and average solutions}

Once a set of solutions has been produced, usually a selection of the best solutions are averaged and then compared against the diffraction data, e.g. by plotting the Phase Retrieval Transfer Function (PRTF)\cite{shapiro_biological_2005, chapman_high-resolution_2006, chushkin_three-dimensional_2014}, which is the ratio of the average calculated amplitude to the observed amplitude, as a function of the resolution ring (a fraction of the sampling frequency of the dataset). The relative frequency at which the PRTF falls below 50\% can be used as an indication of the correlation between the chosen solutions, and of the resolution of the reconstruction.

An alternative approach to averaging consists in computing eigenvectors for the selected solutions. This is done by arranging all solutions in a matrix (by flattening each solution as a one-dimensional row of the matrix), and then computing the singular value decomposition (SVD) of this matrix, which yields a set of orthonormal 'eigen-solutions' \cite{virtanen_scipy_2019}. One property of this decomposition is that it conserves the total squared amplitude of the solutions.

\begin{figure}[ht]
    \centering
    \includegraphics[width=0.7\columnwidth]{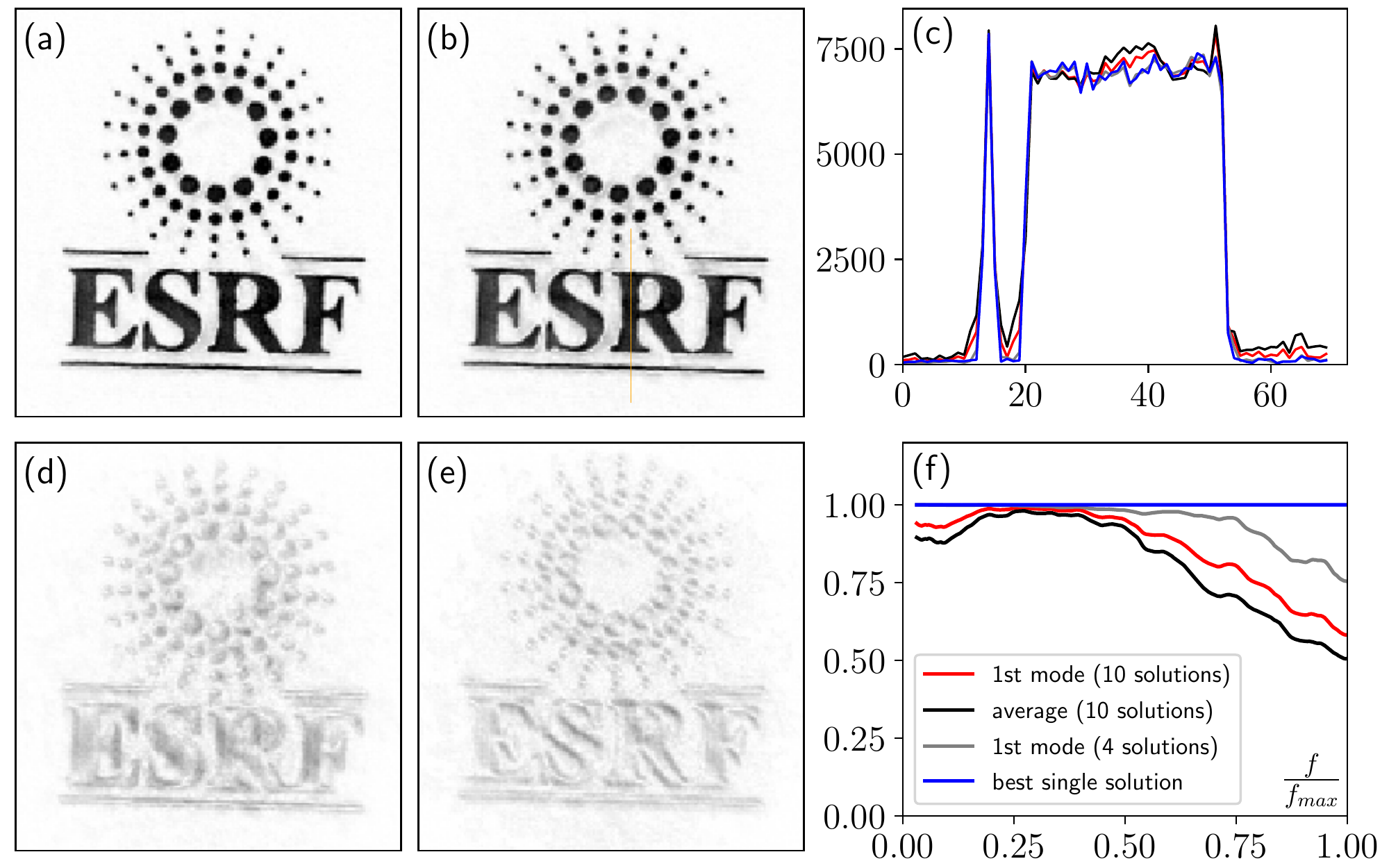}
    \caption{Combination of ten solutions: (a) the first eigen-solution obtained by computing an eigenvalue decomposition (76\% weight), (b) average of the solutions (c) line cut (indicated in orange in (b)), with the first eigen-solution in red, the average solution in black, the best individual solution in blue, and in gray the first eigen-solution when combining only the 4 best solutions (weighting 99\% of the 4 solutions). (d) 2$^{nd}$ and (e)  3$^{rd}$ eigen-solutions in the ten mode decomposition. (f) phase retrieval transfer function computed for the different types of solutions. The solution obtained through eigen-decomposition of the 10 solutions yields a higher correlation at high frequencies, and presents lower background noise levels compared to the average solution. Combining only the 4 best solutions yields a higher PRTF.}
    \label{fig:fig_modes}
\end{figure}

An example of application on the ESRF logo dataset is shown in Fig. \ref{fig:fig_modes}. This  method presents a few advantages compared to averaging: first, it is less sensitive to outlier solutions than averaging (a single outlier would contribute to secondary modes in the eigenvector decomposition), and second, it yields a weight (its overall squared amplitude) for each eigen-solution that can also be used as an indicator of the correlation between solutions, ideally the relative weight of the first (strongest) eigen-solution should be as close to 100\% as possible.

Fig. \ref{fig:fig_modes} shows the result of combining either the 10 or 4 best (lowest $LLK_{free}$) solutions from 50 optimisations, obtained with the same procedure as those in Fig.\ref{fig:fig_llk_evolution}, and after sub-pixel alignment\cite{guizar-sicairos_efficient_2008} and phase matching of the solutions. Note that the 10 best selected solutions had up to 16300 points in the support and a $LLK_{free}$ up to 255, while the 4 best had up to 5100 points with a $LLK_{free}$ lower than 148. The purpose of keeping all 10 solutions is to evaluate the efficiency of combining imperfect solutions, which is often the case in CDI. The shown figures were tested against several generated set of solutions with similar results.

The most intense mode (Fig. \ref{fig:fig_modes}a)) represents 76\% of the 10 solutions, and is similar to the average (Fig. \ref{fig:fig_modes}b)) but with slightly reduced background noise outside the main object (also see the line cut in Fig.\ref{fig:fig_modes}c)). The secondary modes (e.g. Fig. \ref{fig:fig_modes}d) and e)) can be used as an indicator of where the computed solutions have more diversity. The PRTF (Fig. \ref{fig:fig_modes}f)) shows that the first decomposed mode yields a higher resolution than the average solution, but remains inferior to the first mode of only the 4 best solutions (this mode then represents 99\% of the selected 4). Note that when only a subset of close-to-optimal solutions are selected, there is little difference between the first mode and the average (identical PRTF).

\section*{\label{sec:soft}Example application to single cyanobacteria cells}

As the reconstruction relies heavily on the tight support, it is easily obtained for binary objects like the ESRF logo already presented. However in the case of biological specimens (e.g. cells)\cite{miao_imaging_2003, thibault_reconstruction_2006, nelson_high-resolution_2010} the support update often requires some careful optimisation, including hand-picking of the support area \cite{nelson_high-resolution_2010}, as it is difficult to choose a threshold value for the automatic update of the support region using a shrinkwrap-based algorithm \cite{marchesini_x-ray_2003} or it will yield solutions with similar figures-of-merit which must then be sorted out e.g. using clustering analysis \cite{van_der_schot_imaging_2015}.

To test the usefulness of free log-likelihood for biological samples, we used the Coherent X-ray imaging Database \cite{maia_coherent_2012} dataset \#16\cite{schot_imaging_2015}, which includes the original diffraction data and the published experimental parameters\cite{van_der_schot_imaging_2015}.

\begin{figure}[ht]
    \centering
    \includegraphics[width=0.7\columnwidth]{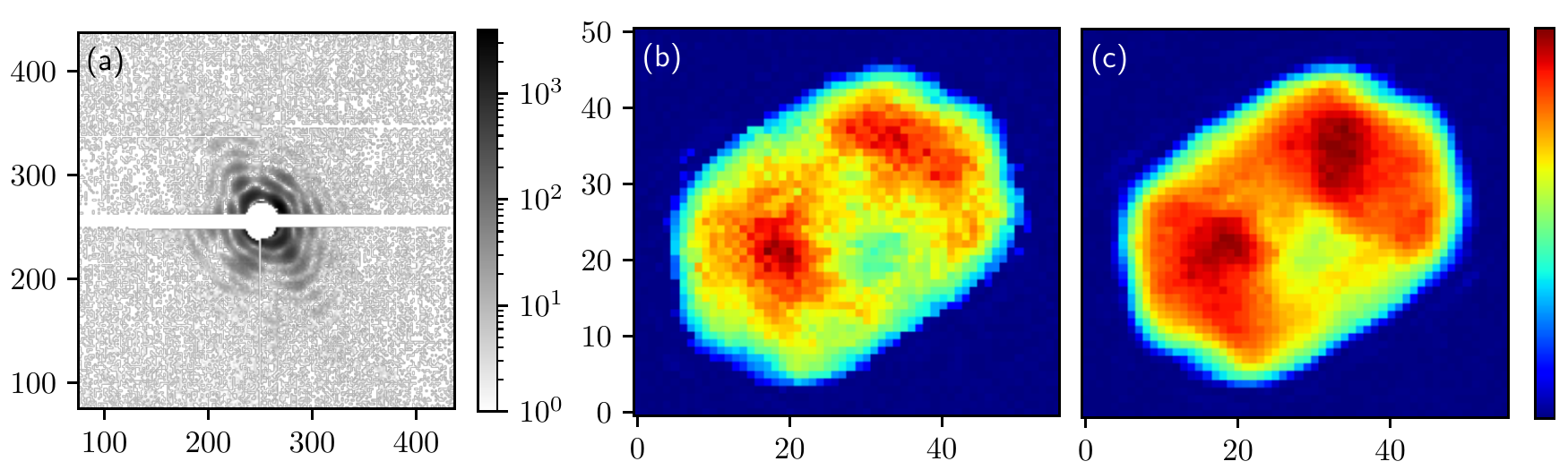}
    \caption{Example data analysis on coherent diffraction imaging of single cyanobacteria cells: (a) coherent diffraction pattern, (b) amplitude of the best individual reconstruction (with the lowest free log-likelihood) and (c) of the first eigen-mode obtained by decomposing the 10 best results among the 100 optimisations, representing 97\% of the modes. This can be compared to Fig. 5g and 6 in the original publication \cite{van_der_schot_imaging_2015}, which uses the same linear colour scale.}
    \label{fig:fig_cell}
\end{figure}

We used the following procedure, with the goal to test unsupervised support and object optimisation:
\begin{enumerate}
    \item Run 100 independent reconstructions with the following parameters: start from a large support 200 pixels in diameter, with a random object. First perform 2000 cycles of HIO with a positivity constraint and a detwinning procedure after 1000 cycles. Then proceed without the positivity constraint, with 2000 HIO cycles and then 2000 RAAR cycles. Finish with 200 ER cycles. The support was updated every 20 cycles, with a fixed relative threshold randomly chosen between 0.1 and 0.4.
    \item Select the best 10 reconstructions based on the free log-likelihood, and perform an eigen-solution analysis. Note that during this analysis, the mask of free pixels was kept fixed (contrary to what is shown in Fig. \ref{fig:fig_llk_evolution}) for all optimisations, in order to ensure the consistency of the free log-likelihood figure of merit.
\end{enumerate}

The result of this optimisation is shown in Fig. \ref{fig:fig_cell}. The results can be compared to the original publication, where 400 optimisations where performed for each dataset, and clustering analysis was performed to detect outliers with similar figures of merit (Fig. 6 in \cite{van_der_schot_imaging_2015}). Our results are similar but require a less intensive computational approach.

\section*{\label{sec:conclusion}Conclusion}
In this article we have shown that using a free log-likelihood figure-of-merit allows one to evaluate solutions from CDI optimisations in a unbiased manner, despite the lack of \textit{a priori} knowledge of the object support size and shape. Moreover using an orthonormal mode decomposition of the best solutions yields a better solution less prone to outlier results compared to the usual averaging approach.

 The main advantage of this method is that it is completely generic, as it does not rely on any \textit{a priori} knowledge on the sample (complex-valued or real-valued object, homogeneity of density or phase, etc.), and can thus be used for \textit{unsupervised} phasing. Moreover this approach has a very low computational cost and can easily be implemented. This should be particularly relevant for high data throughput approaches which are now being developed with X-ray free electron laser and brighter synchrotron sources.

\bibliography{Article-FreeLLK}

\section*{Acknowledgements}
The authors would like to acknowledge Federico Zontone for help with the CDI measurements, and Pierre Thibault for a discussion on the mode decomposition as implemented in Ptypy \cite{enders_computational_2016}, on which was based the eigen-decomposition for CDI proposed in this article.

\section*{Author contributions statement}
VFN proposed the algorithms, contributed the PyNX library for computations, and wrote the main manuscript text. SL and YC provided critical feedback on the procedure and reviewed the manuscript. YC collected the experimental data for figures 1-3.

\section*{Additional information}
The authors declare no competing interests.

\newpage
\part*{Supplemental information}
\section*{Free log-likelihood curve vs free pixels island size}
\renewcommand{\figurename}{Suppl. Fig.}

As indicated in the main text, the free log-likelihood is calculated by setting aside $\approx$ 5\% of the observed diffraction data in a 'free' set of pixels, which are grouped as islands of radius 3 pixels - to make sure that correlations between neighbouring pixels does not create a strong relationship between the working set and the 'free' set.”

In the following figure \ref{fig:supplFig5} we have performed the same calculations and plot as for figure \ref{fig:fig_llk_evolution} of the article, but by changing the island size with a radius varying from 0 (isolated pixels) to 6. The radius is indicated for each figure in the legend.
	
As can be seen in these figures, the free log-likelihood is not discriminating enough for the small island’s radii (0 and 1), as incorrect solutions with a large support still can yield small $LLK_{free}$ values. This is particularly true when the radius is equal to zero, as the $LLK_{free}$ has the same tendency as the normal log-likelihood, i.e. it is decreasing with an increasing number of pixels in the object support. This confirms that using islands with a sufficient radius is necessary to yield a discriminating figure of merit, in order to obtain a sufficient  
\newpage

\begin{figure}[H]
    \centering
    \begin{tabular}{c|c}
     \includegraphics[width=0.4\columnwidth]{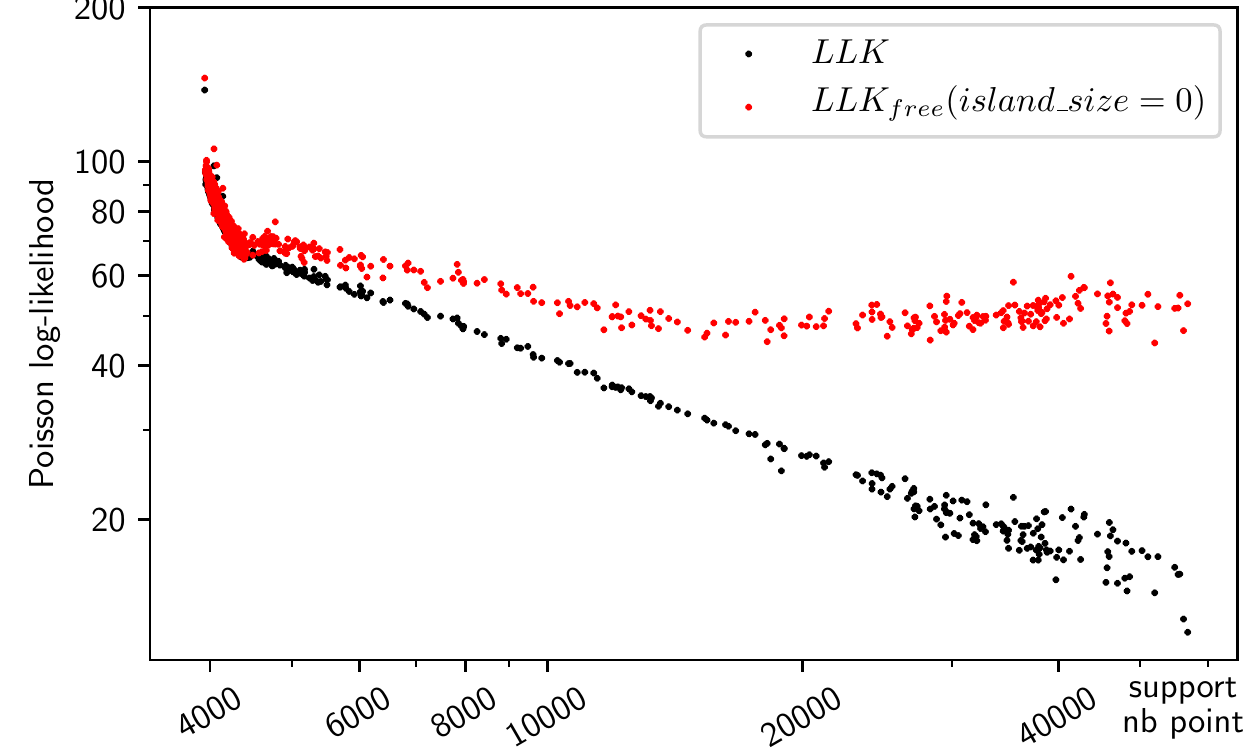}   &  \includegraphics[width=0.4\columnwidth]{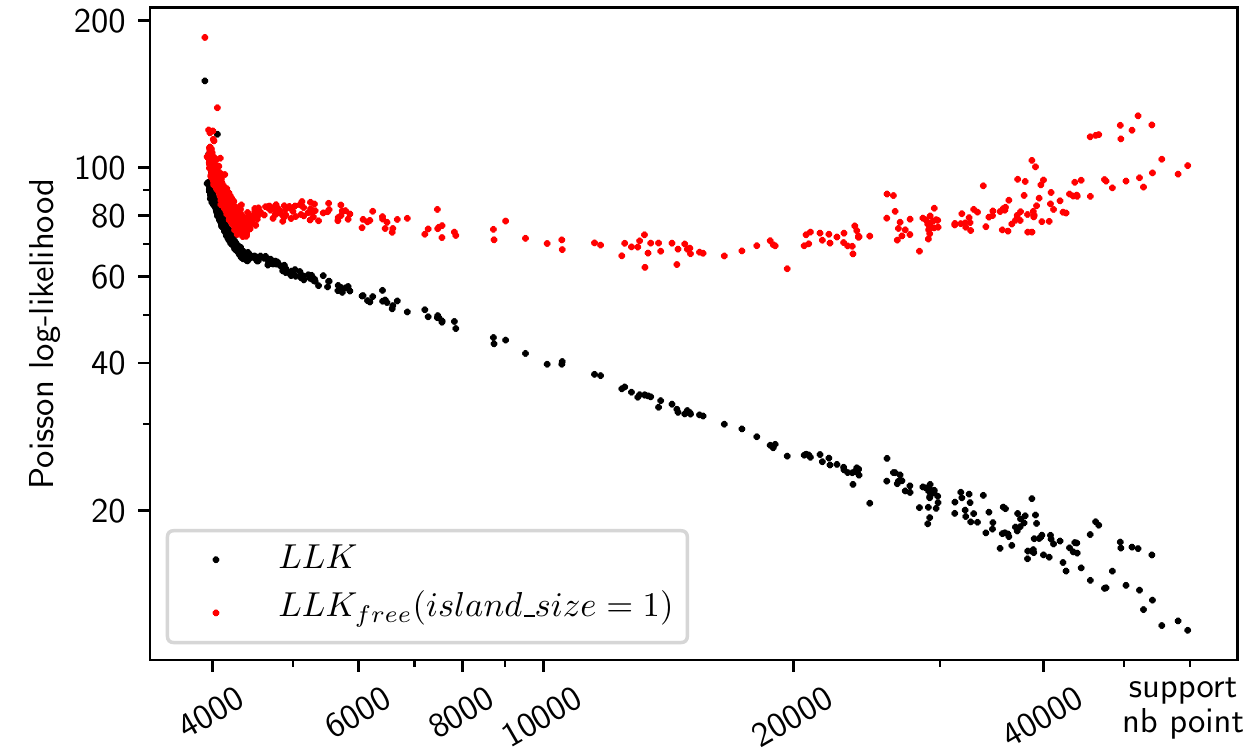}\\
     \includegraphics[width=0.4\columnwidth]{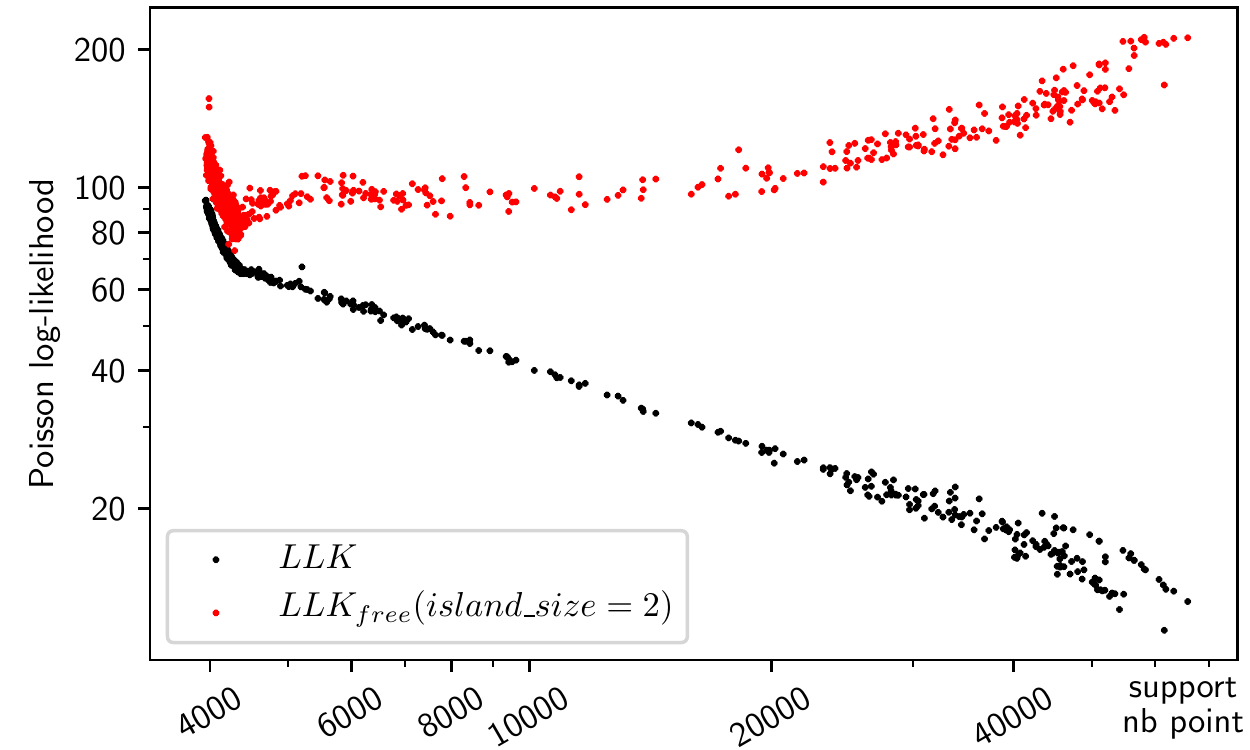}   &  \includegraphics[width=0.4\columnwidth]{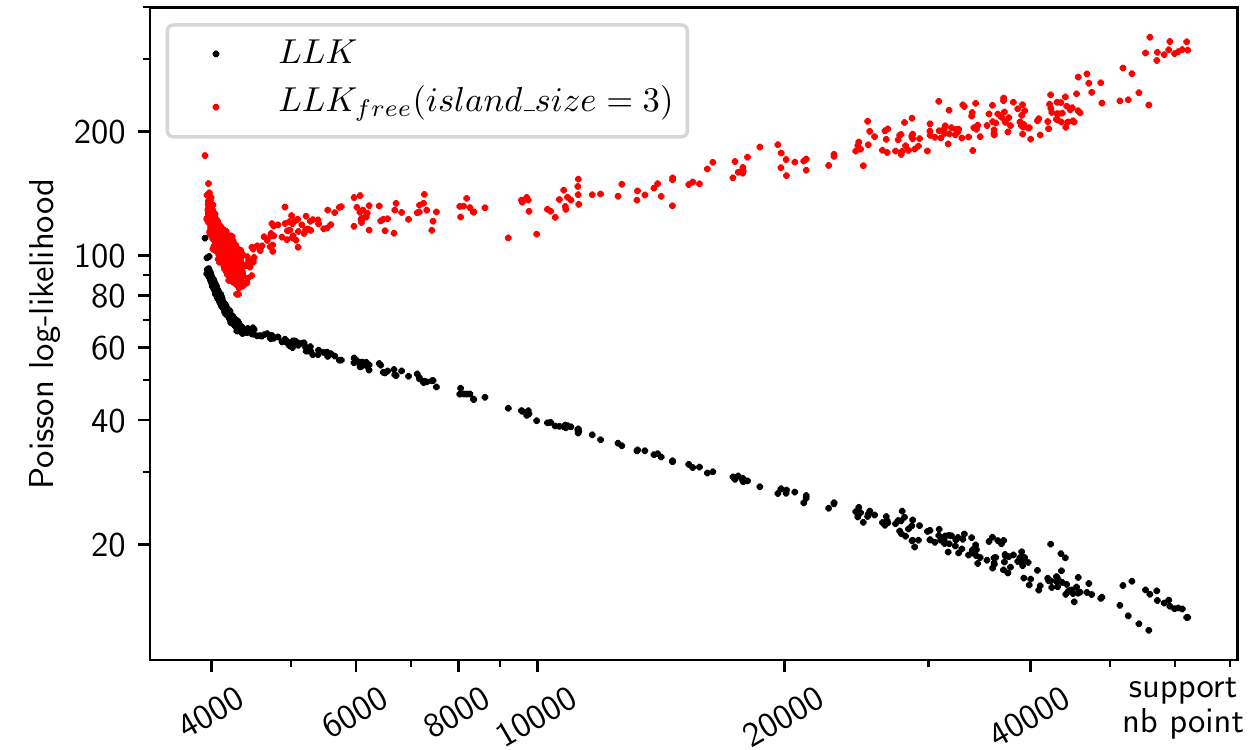}\\
     \includegraphics[width=0.4\columnwidth]{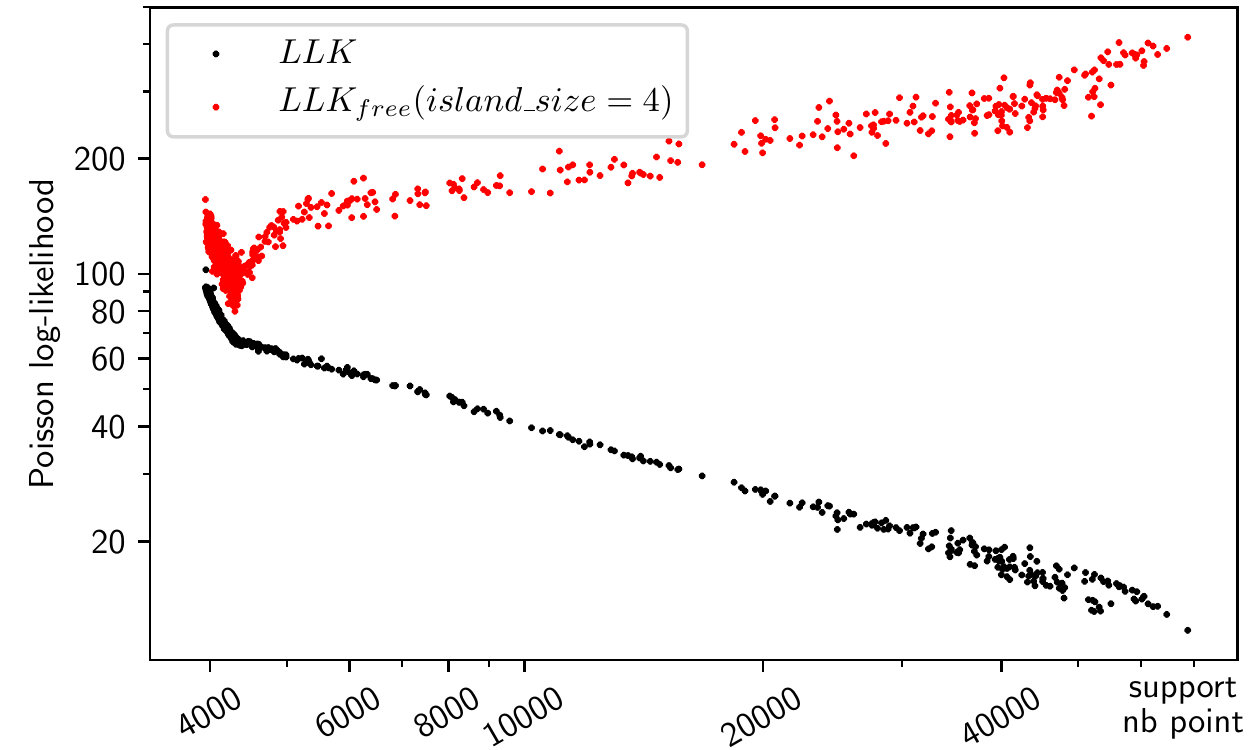}   &  \includegraphics[width=0.4\columnwidth]{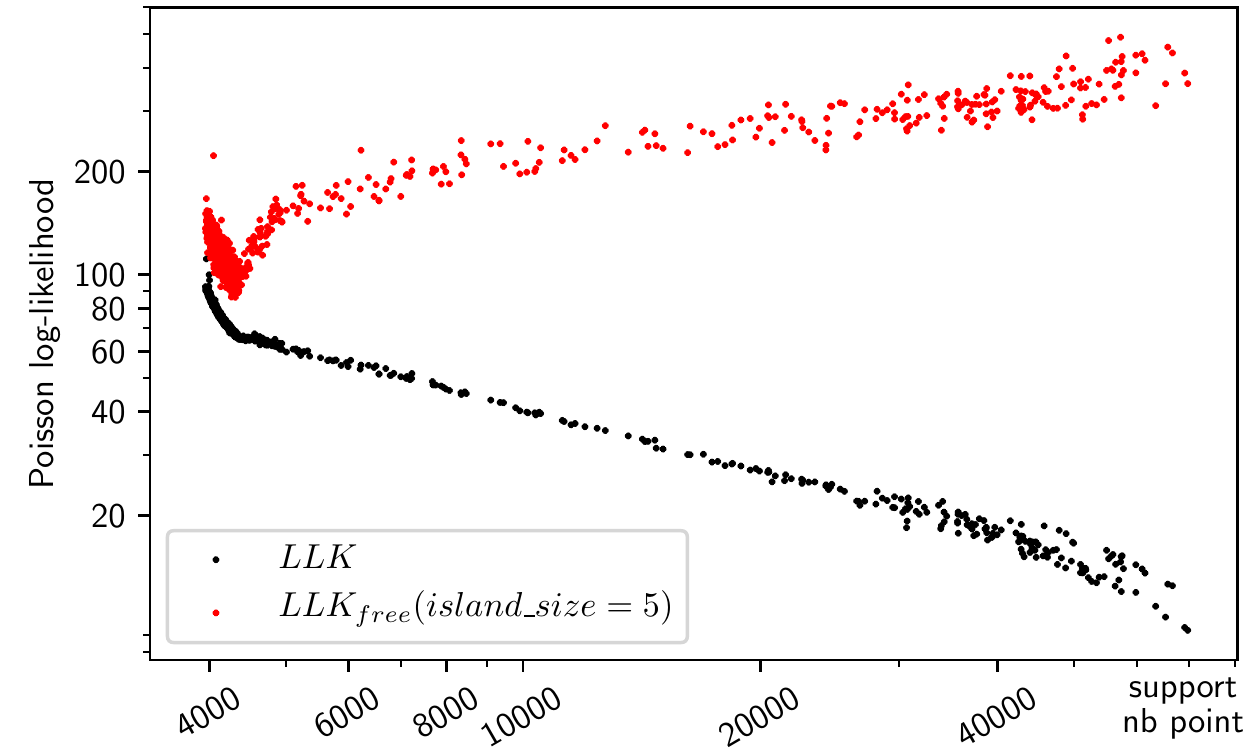}\\
     \includegraphics[width=0.4\columnwidth]{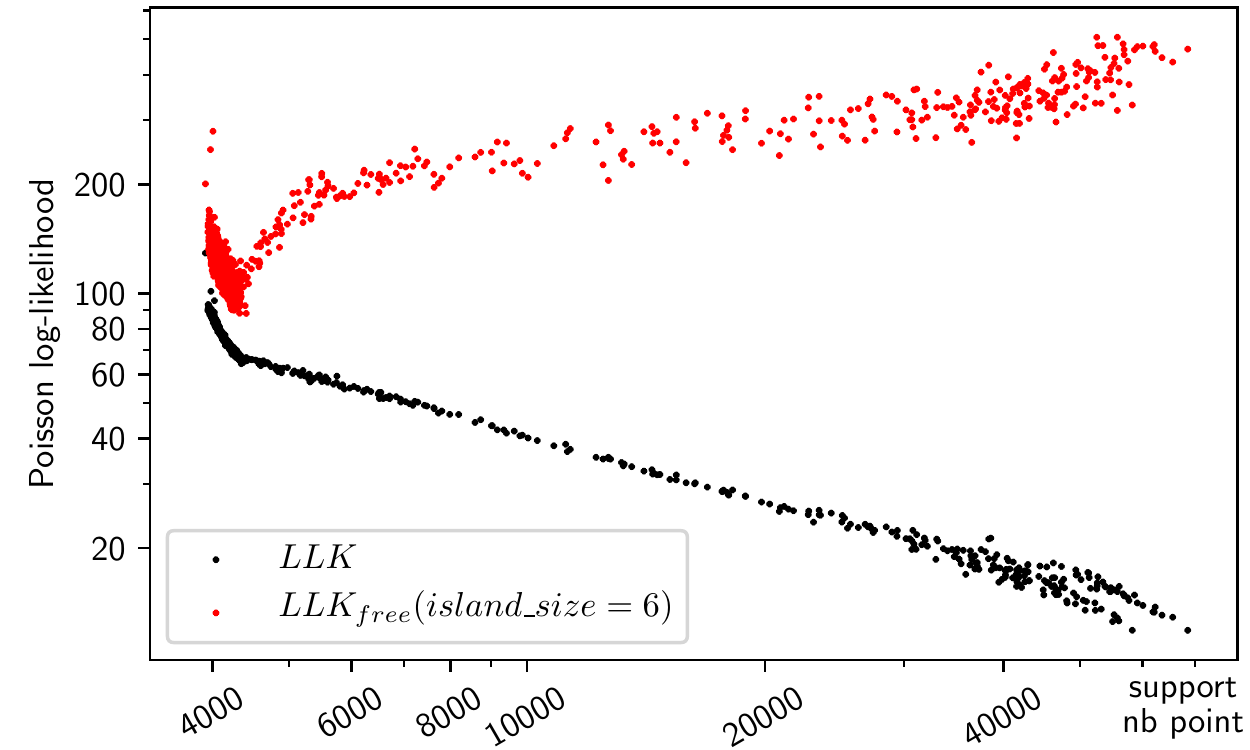}   &  \\ 
    \end{tabular}
    
    \caption{scatter plots of log-likelihood (black) and free log-likelihood (red) vs the number of points in the support, obtained by generating 1000 solutions with random starting objects and a different threshold. Each plot corresponds to a different calculation of LLKfree, with ‘free’ pixels grouped as islands with a radius (indicated in the legend) varying from 0 to 6 pixels. Each plot also corresponds to 1000 independent solutions with different random start and a random threshold}
    \label{fig:supplFig5}
\end{figure}

\newpage
\section*{Free log-likelihood curve for the cyanobacteria data}
The curve was generated similarly to figures \ref{fig:fig_llk_evolution}{} and (suppl) \ref{fig:supplFig5}, but for the cyanobacteria dataset presented in figure 4. The overall behaviour is similar to figure 2, with the normal log-likelihood decreasing with increasing number of points in the support, whereas the free log-likelihood presents a minimum around the ideal support size. 
	
In this particular case the minimum is less pronounced than in Fig.2, due to the faceted shape of the bacteria which allows relatively easy convergence of the algorithm towards a correct shape.

\begin{figure}[H]
    \centering
    \includegraphics[width=0.7\columnwidth]{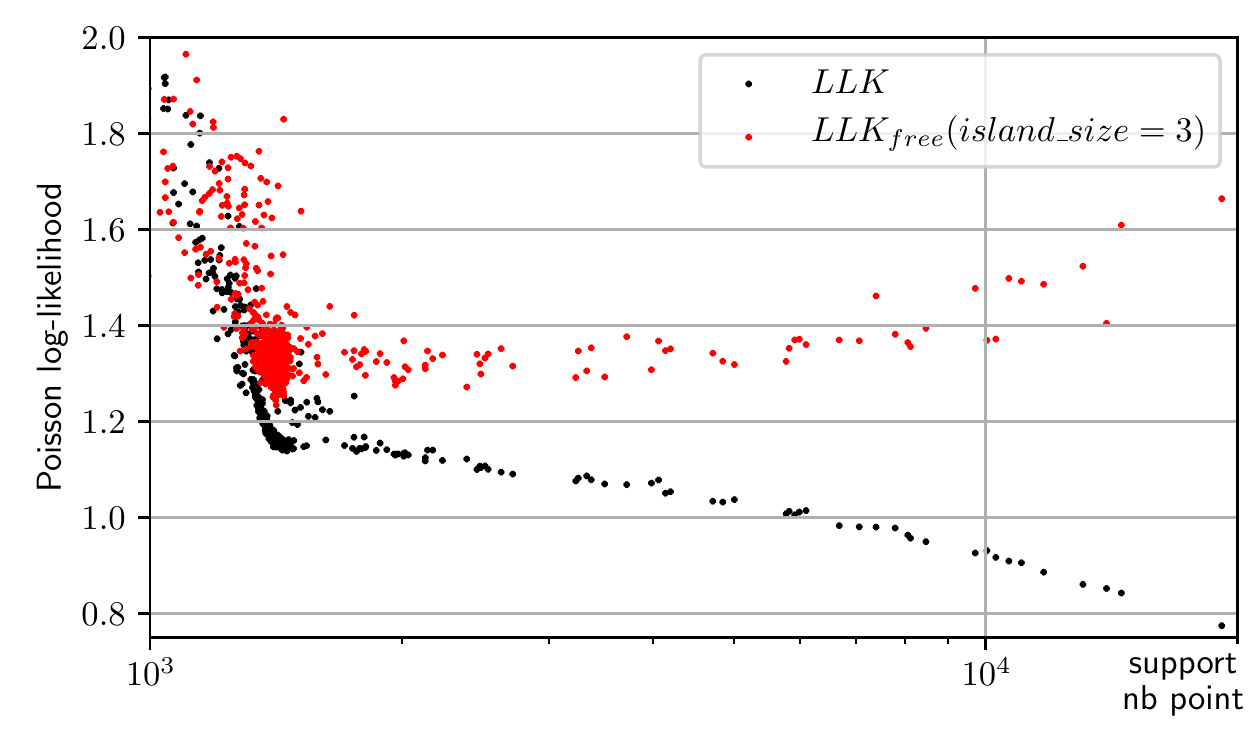}
    \caption{scatter plots of log-likelihood (black) and free log-likelihood (red) vs the number of points in the support, obtained by generating 1000 solutions with random starting objects and a different threshold.}
    \label{fig:supplFig6}
\end{figure}
\end{document}